\PassOptionsToPackage{svgnames,table}{xcolor}
\documentclass[sigconf, authorversion]{acmart}

\AtBeginDocument{%
  }

\copyrightyear{2025}
\acmYear{2025}
\acmConference[WWW Companion '25]{Companion Proceedings of the ACM Web Conference 2025}{April 28-May 2, 2025}{Sydney, NSW, Australia}
\acmBooktitle{Companion Proceedings of the ACM Web Conference 2025 (WWW Companion '25), April 28-May 2, 2025, Sydney, NSW, Australia}
\acmDOI{10.1145/3701716.3715597}
\acmISBN{979-8-4007-1331-6/2025/04}
\usepackage{multirow}
\usepackage{graphicx}

\usepackage{pifont}    
\usepackage{caption}   
\usepackage[table]{xcolor}

\usepackage{multirow}
\usepackage{tabularx}
\usepackage{makecell}

\usepackage{booktabs}
\usepackage{threeparttable}
\usepackage{microtype}

\usepackage{enumitem}
\usepackage{setspace}

\begin{document}

\title{Generating Privacy-Preserving Personalized Advice with Zero-Knowledge Proofs and LLMs}

\author{Hiroki Watanabe}

\email{watanabe.hiroki@jri.co.jp}
\affiliation{%
  \institution{The Japan Research Institute, Ltd.}
  \state{Tokyo}
  \country{Japan}
}

\author{Motonobu Uchikoshi}

\email{uchikoshi.motonobu@jri.co.jp}
\affiliation{%
  \institution{The Japan Research Institute, Ltd.}
  \state{Tokyo}
  \country{Japan}
}

\renewcommand{\shortauthors}{Hiroki Watanabe and Motonobu Uchikoshi}
\raggedbottom
\begin{abstract}
Large language models (LLMs) are increasingly utilized in domains such as finance, healthcare, and interpersonal relationships to provide advice tailored to user traits and contexts. However, this personalization often relies on sensitive data, raising critical privacy concerns and necessitating data minimization. To address these challenges, we propose a framework that integrates zero-knowledge proof (ZKP) technology, specifically zkVM, with LLM-based chatbots. This integration enables privacy-preserving data sharing by verifying user traits without disclosing sensitive information. Our research introduces both an architecture and a prompting strategy for this approach. Through empirical evaluation, we clarify the current constraints and performance limitations of both zkVM and the proposed prompting strategy, thereby demonstrating their practical feasibility in real-world scenarios.
\end{abstract}


\maketitle

\section{Introduction}
Large language models (LLMs) are increasingly providing advice through chatbots in fields such as finance, healthcare, and interpersonal relationships. As LLM-generated advice shapes decision-making, incorporating user characteristics and contexts is crucial for enhancing its effectiveness\cite{wester2024exploring,furumai-etal-2024-zero,bhattacharjee2024understanding}. 

However, tailoring advice to individual circumstances raises significant privacy concerns. For example, financial advice may require sensitive information like income and assets, while medical advice necessitates personal details such as age and medical history to ensure accuracy. Many chatbot platforms outline privacy policies that detail data collection and usage, allowing users to opt out or request data deletion. Despite this, platforms often collect more data than necessary, underscoring the importance of the data minimization principle. This principle is emphasized in the European General Data Protection Regulation (GDPR) and reflected in the proposed American Data Privacy and Protection Act (ADPPA), currently under legislative review.

One promising technology for achieving data minimization is zero-knowledge proofs (ZKP), which have seen significant advancements in the blockchain field. ZKP frameworks, such as zk-SNARKs, efficiently prove the correctness of program execution as part of verifiable computation, without revealing sensitive data \cite{thaler2022proofs}. This technology is particularly useful in blockchain applications, where repeated transaction verification is necessary, but its potential is not limited to that domain. For instance, it can enable the provision of verifiable user characteristics and circumstances to LLM platforms without disclosing detailed personal information.

Despite the promise of this technology, research on its application in LLM-based chatbots remains limited. One prior study by Cai et al.\cite{Cai2023hcpp} explored the use of ZKP in healthcare chatbots. Its application, however, was restricted to relatively few data points and relied on basic logic, such as verifying if a value satisfied a range proof. This highlights the limited flexibility of current zk-frameworks like zk-SNARKs. Recent advancements, such as zkVM, are expected to address these limitations for practical ZKP applications. Yet, evaluations of zkVM’s capabilities and performance remain limited, leaving its potential for real-world implementation largely unexplored. Moreover, effectively using verifiable user traits generated through ZKP for advice generation using LLMs remains a research challenge.

The goal of this research is to develop a new framework for LLM-based chatbots that delivers more personalized and appropriate advice while protecting user privacy through the use of zero-knowledge proof technology. To achieve this, we address the following research questions:
\begin{spacing}{0.8}
\begin{itemize}[itemsep=1.1mm, topsep=1.72mm, leftmargin=3mm]
    \item To the best of our knowledge, this study is the first to explore the potential of combining a modern ZKP framework, zkVM, with an LLM-based advisor for privacy-preserving applications.
    \item We propose an architecture and a prompting strategy that integrate zkVM and LLMs, enabling secure verification of user traits without disclosure and facilitating consistent, privacy-preserving advice generation.
    \item Through empirical evaluation, we clarify the current constraints and performance limitations of both the zkVM and the proposed prompting strategy, thereby demonstrating their practical feasibility.
\end{itemize}

\end{spacing}
\section{Zero-Knowledge Proofs}
Zero-Knowledge Proofs (ZKP) are cryptographic techniques that enable one to prove possession of specific information without revealing the information itself \cite{thaler2022proofs}. In recent years, ZKPs have rapidly evolved, particularly with the development of practical zk-SNARKs (Succinct Non-Interactive Arguments of Knowledge)\cite{parno2013pinocchio,groth2016size}, which have been widely adopted in the blockchain domain for applications such as protecting transaction privacy in cryptocurrencies and compressing data using rollup techniques. This technology is expected to be applied across various fields beyond blockchain due to its ability to verify computations efficiently and privately, holding significant potential for broader adoption.

\subsection{Related Work: Combining LLMs and ZKPs}
Efforts to combine LLMs with ZKP technology are in their early stages. zkLLM \cite{sun2024zkllm} aims to enable verification of the entire inference process of an LLM using ZKP. In this framework, the model owner commits to their model parameters, and the commitment is used to prove that the inference process has been executed correctly using those parameters. Meanwhile, HuRef \cite{zeng2024huref} focuses on copyright protection for LLMs by proposing “human-readable fingerprints” to identify LLMs. ZKP is used to demonstrate that the claimed LLM parameters were used to generate the fingerprints accurately. While zkLLM and HuRef primarily focus on model verification using ZKP technology, Singh’s study \cite{singh2024enhancing} explores potential applications related to the data input to LLMs, such as user authentication, prompt analysis, source data verification, and source data relevance filtering. In this context, our use case specifically emphasizes the protection of user-provided data. A related example is presented by Cai et al. \cite{Cai2023hcpp}, who demonstrated a chatbot in the healthcare domain using ZKP to protect sensitive user information. However, their ZKP application was limited to four data points—age, height, weight, and medical history—and relied on a simple implementation, such as range proof verification, resulting in a relatively small-scale application. Additionally, the design of prompts to effectively utilize ZKP-generated outputs within LLM chatbots remains an unexplored area. This field is evolving rapidly and highlights a promising direction for privacy-focused LLM applications. However, further advancements require more practical implementations and detailed evaluations tailored to specific and complex use cases.

\begin{table*}[tbp]
\centering
\small
\setlength{\tabcolsep}{3pt}
\renewcommand{\arraystretch}{1}
\caption{Comparative Analysis of Popular zkVM Projects}
\label{tab:zkvm_comparison}
\resizebox{\textwidth}{!}{
\begin{threeparttable}
\begin{tabular}{lllccl}
\toprule
\textbf{Project} & \textbf{Supported Languages} & \textbf{ISA (Instruction Set Architecture)} & \textbf{Proof System} & \textbf{Crate/Package Import Support} & \textbf{Private Inputs} \\
\midrule
CairoVM (Starkware) & Cairo & Proprietary Architecture & STARK & No & Yes \\
MidenVM (Polygon Miden) & Miden Assembly & Proprietary Architecture & STARK & No & Yes \\
Valida & C, Rust & Proprietary Architecture & SNARK & No & N/A\tnote{*} \\
Nexus & Rust (no std) & General Architecture (RISC-V) & SNARK & No & Yes \\
zkWASM & C (no std), Rust (no std) & General Architecture (WASM) & SNARK & Yes & Yes \\
Jolt (a16z) & Rust & General Architecture (RISC-V) & SNARK & Yes & N/A\tnote{*} \\
SP1 (Succinct Labs) & Rust & General Architecture (RISC-V) & STARK/SNARK & Yes & Yes \\
RiscZero & Rust & General Architecture (RISC-V) & STARK/SNARK & Yes & Yes \\
\bottomrule
\end{tabular}
\begin{tablenotes}
\footnotesize
\item [*] N/A indicates features that are currently not implemented as of this writing but may be added in future updates.

\end{tablenotes}
\end{threeparttable}
}
\end{table*}

\section{Architecture and Methodology}
\begin{figure}[tbp]
\centering
\includegraphics[width=0.95\columnwidth]{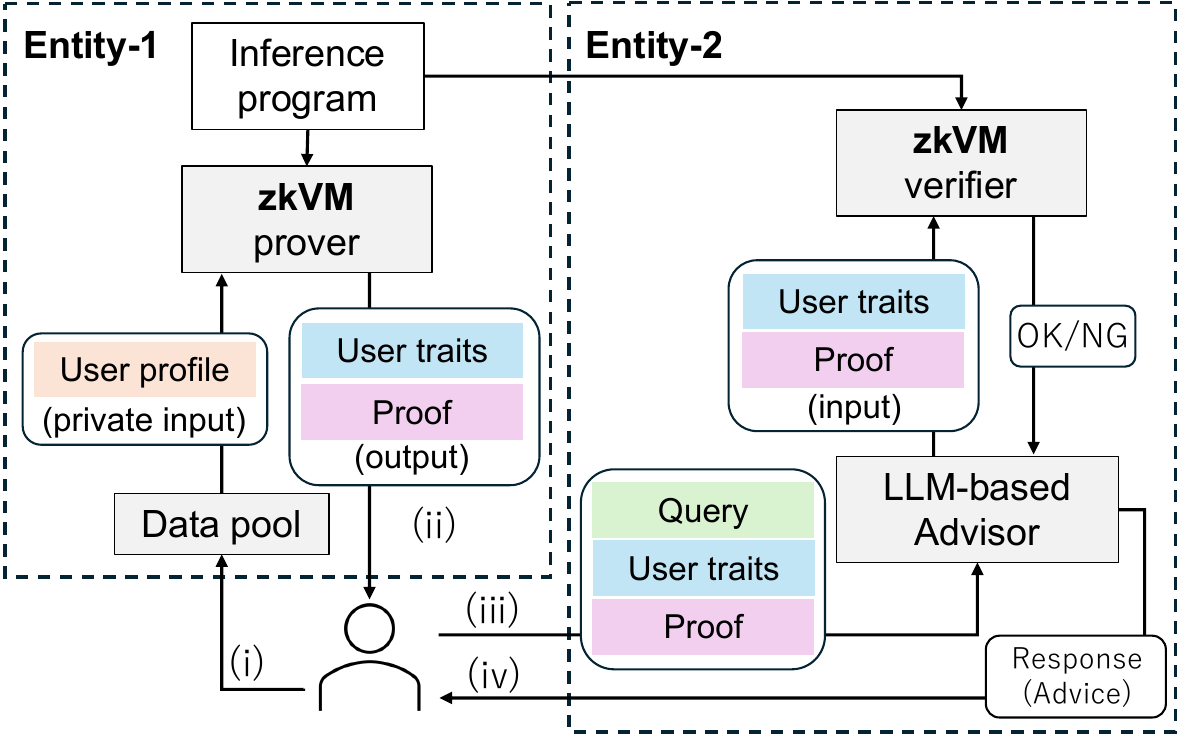}
\caption{Overview of the Architecture for LLM-based Advisory System with Zero-Knowledge Proofs}
\label{fig:fig1}
\end{figure}

The architecture we propose is operated by two independent entities, as shown in Figure \ref{fig:fig1}. Entity 1 can be assumed to be an institution that holds user data, such as a financial institution or a healthcare provider. Entity 2, on the other hand, is envisioned as a cloud-based service that utilizes LLMs to provide advice to users. These two entities operate independently, and the sharing of personal data between them is restricted.
(i) The user requests Entity 1 to infer abstracted user traits based on their specific profile, (ii) receiving the results along with a zero-knowledge proof. The proof and user traits are then (iii) provided to Entity 2 alongside a query, and (iv) the user receives feedback from an LLM-based advisor system or agent.

A key feature of this architecture is that the output from Entity 1 contains both the inferred user traits and the proof verifying those traits. Entity 2 can verify the proof to ensure that the user traits were computed according to a predefined algorithm.

\subsection{Introduction of zkVM}
In previous studies \cite{singh2024enhancing, Cai2023hcpp}, the approach of constructing arithmetic circuits from programs designed for ZKPs and transforming them into polynomials for proof generation was adopted, based on the zk-SNARKs. While this method offers advantages in terms of proof size and proof generation time, it presents challenges in requiring code optimized for arithmetic circuits. Additionally, developers must use domain-specific languages (DSLs) or specialized libraries, leading to high learning costs. Furthermore, when implementing complex algorithms, developers often seek to leverage existing software libraries and packages commonly utilized in high-level programming languages; however, the necessity to optimize code for arithmetic circuits frequently constrains their ability to do so.

To address these challenges, our study adopts zkVM (Virtual Machine), a modern zero-knowledge proof architecture. zkVM enables the execution of entire programs as zero-knowledge proofs, characterized by its flexibility and versatility. Specifically, instead of directly converting the program into arithmetic circuits, zkVM executes the program and proves the consistency of its execution trace (such as memory and register state transitions). This approach allows for quick verification that a program executed on zkVM produces the same output given the same input, without needing to rerun the program. Additionally, zkVM leverages zero-knowledge properties, enabling some parts of the input to remain private. One of zkVM's main advantages is that it can utilize existing programs with minimal modification, allowing developers to integrate zero-knowledge proof technology using familiar high-level languages and libraries. This significantly reduces the learning curve compared to traditional zk-SNARK architectures that require extensive optimization for arithmetic circuits and often necessitate the use of domain-specific languages or specialized libraries.

To utilize zkVMs in our research, we investigated several popular zkVM projects, as shown in Table~\ref{tab:zkvm_comparison}. Our selection criteria focus on implementations that do not rely on blockchain or smart contracts and can run in a local environment. While numerous zkVM projects exist, considering factors such as support for familiar high-level languages, package imports, and input privacy, RiscZero\cite{bruestle2023risc} and SP1\cite{sp1book} emerge as practical options at this time.

\subsection{Prompt Strategy}

With the introduction of the zkVM, we can query the LLM advisor with user traits and proofs, as shown in Figure~\ref{fig:fig1}. We propose the following prompting strategy.

Let \( L \) represent text containing user traits, partitioned into \( \mathcal{D} = \{d_0, d_1\} \). Here, \( d_0 \) denotes unverifiable exploratory traits, and \( d_1 \) represents traits verifiable through zero-knowledge proof (ZKP) schemes, which have higher proof reliability. Verifiable traits \( d_1 \) are supported by objective evidence or logical constraints, enabling efficient validation using ZKP technology. Based on this partition, we define a context set \( C = \{c_0, c_1, \dots, c_n\} \), where each \( c_i \in C \) is generated by varying the emphasis on \( d_0 \) and \( d_1 \).

For a user query \( Q \), the LLM generates a response divided into two parts: a proposed answer \( A_{\text{prop}} \) and an explanation \( A_{\text{exp}} \). Each part is generated using distinct instructions and their respective contexts, \( c_{\text{prop}} \) and \( c_{\text{exp}} \). These contexts may either be same or tailored to address different aspects of the query and are selected from the context set \( C \). The responses are computed as follows:
\[
A_{\text{prop}} = \text{LLM}(Q, I_{\text{prop}}, c_{\text{prop}}), \quad 
A_{\text{exp}} = \text{LLM}(Q, A_{\text{prop}}, I_{\text{exp}}, c_{\text{exp}}).
\]

The objective of this strategy is to create personalized and consistent responses by leveraging both unverifiable and verifiable user traits. 

\section{Experiments}
\subsection{Evaluation of zkVM}
The evaluation of the zkVM focuses on measuring the proof generation and verification times of the created ZKP application to assess its feasibility. While existing benchmarks \cite{risczero_reports} primarily emphasize basic cryptographic operations, our study shifts the focus to real-world applications, such as user categorization logic designed to infer user traits.

\subsubsection{Application}

One of the central questions of this study is whether real-world inference applications can be replicated as ZKP applications. As a first step, we chose to focus on rule-based inference applications, specifically a web application published by the Japanese Bankers Association\footnote{Available at \url{https://www.zenginkyo.or.jp/article/tag-c/diagnosis/risktest/}. (Japanese)}. This application categorizes users into one of four risk tolerance categories based on their responses to 10 financial questions (e.g., whether they have a mortgage or a retirement plan), which are used to infer user traits. Each question offers three response options, and the application uses rule-based categorization logic to classify users into the following categories: (i) Conservative, (ii) Steady Growth, (iii) Balanced, or (iv) Aggressive Investment.

We recreated this application using general-purpose Rust code and implemented it as a ZKP application on a zkVM. The application performs the following operations: parsing private inputs in JSON format, concatenating question strings, applying SHA-256 hashing, scoring responses, and classifying users into predefined categories. To evaluate the ZKP application and the zkVM frameworks—RiscZero and SP1, as described in Section 3.1—we prepared a dataset by randomly sampling 40 user profiles from all possible response scenarios (a total of 59,049 combinations), ensuring an even distribution across the four risk tolerance categories. Each profile is unique, but the same inference logic is applied to all users.

\begin{table}[tbp]
\centering
\small
\setlength{\tabcolsep}{3pt}
\caption{Benchmark Results of RiscZero and SP1}
\label{tab:benchmark_results}
\resizebox{\columnwidth}{!}{
\begin{threeparttable}
\begin{tabular}{llcc}
\toprule
\multirow{2}{*}{\textbf{Configuration}} & \multirow{2}{*}{\textbf{Metric}} & \textbf{RiscZero} & \textbf{SP1} \\
 & & \textbf{(v1.1.2)} & \textbf{(v3.4.0)} \\
\midrule
\multirow{2}{*}{16 vCPUs, 64GB RAM} & Proof Generation Time & 67.80\,s & 67.21\,s \\
 & Verification Time & 0.02\,s & 1.28\,s \\
\midrule
\multirow{2}{*}{32 vCPUs, 128GB RAM} & Proof Generation Time & 51.80\,s & 57.02\,s \\
 & Verification Time & 0.02\,s & 1.21\,s \\
\midrule
\multirow{2}{*}{AVX2 on 32 vCPU, 128GB RAM} & Proof Generation Time & N/A & \textbf{47.71\,s} \\
 & Verification Time & N/A & \textbf{0.89\,s} \\
\midrule
\multirow{2}{*}{\makecell[l]{CUDA on 1x NVIDIA A100 GPU\\ (24 vCPU, 220GB RAM)}} & Proof Generation Time & \textbf{1.45\,s} & N/A\tnote{†} \\
 & Verification Time & \textbf{0.02\,s} & N/A\tnote{†} \\
\bottomrule
\end{tabular}
\begin{tablenotes}
\footnotesize
\item [†] SP1’s CUDA implementation is still in beta and failed to execute, marked as N/A.
\end{tablenotes}
\end{threeparttable}
}
\end{table}

\subsubsection{Benchmark}
Table~\ref{tab:benchmark_results} summarizes the benchmark results for the zkVM frameworks RiscZero (v1.1.2) and SP1 (v3.4.0) when executing the ZKP application. Zero-knowledge proof generation is known to incur significantly greater overhead compared to verification, and in our inference application, CPU-based proof generation required tens of seconds. On the other hand, real-time applications such as LLM-based chatbots prioritize immediacy, making verification time a more critical factor. The benchmark results indicate that this requirement is effectively met. Furthermore, RiscZero’s substantial performance improvements in proof generation through GPU utilization highlight its potential to enhance usability for user-driven proof requests.

\subsection{Evaluation of Prompt Strategy}

The proposed prompt strategy (described in Section 3.2) aims to generate diverse response variations by emphasizing either unverifiable user traits (\( d_0 \)) or verifiable user traits (\( d_1 \)). This evaluation addresses two key questions: (1) Can the strategy select proposed actions (\( A_{\text{prop}} \)) aligned with the emphasized traits? and (2) Can the strategy generate consistent explanations (\( A_{\text{exp}} \)) for the selected proposals?

\begin{table}[tbp]
\centering
\Large
\setlength{\tabcolsep}{0pt}
\renewcommand{\arraystretch}{1.1}
\caption{Examples of predefined options for generating \( A_{\text{prop}} \) in the healthcare domain.}
\label{tab:positivity_negativity_scores}
\resizebox{\columnwidth}{!}{%
\begin{threeparttable}
\begin{tabular}{>{\Large}p{14.2cm}c}
\toprule
\textbf{Suggested Actions for Headache} & {\textbf{Score}} \\
\midrule
Rest and hydrate to relieve headache, but consider a check-up if persistent. & +2 \\
Take a break for headache relief, with occasional check-ups to catch underlying issues early. & +1 \\
Balance rest for headaches and regular screenings to maintain health. & 0 \\
Opt for check-ups, but allow for headaches due to fatigue. & -1 \\
Prioritize medical check-ups to rule out serious causes, even if the headache feels like fatigue. & -2 \\
\bottomrule
\end{tabular}
\end{threeparttable}
}
\end{table}

\begin{figure}[tbp]
\centering
\includegraphics[width=\columnwidth]{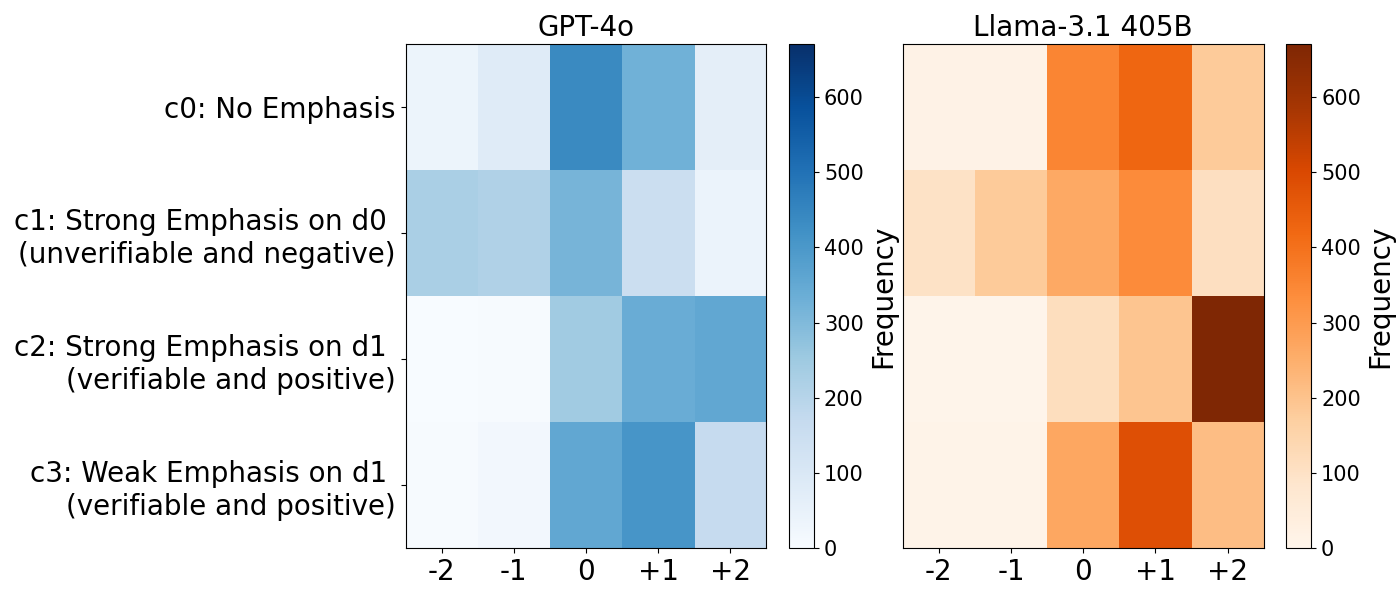}

\caption{Distribution of \( A_{\text{prop}} \) Scores Across Contexts (\( c_i \))}

\label{fig:fig2}
\end{figure}

\subsubsection{Task and Dataset}

This task involves querying a domain-specific LLM advisor with the question, “What action should the user take?” The LLM advisor utilizes GPT-4o, accessed via OpenAI API, and Llama-3.1 405B, accessed via the Amazon Bedrock API, to generate two outputs: \( A_{\text{prop}} \) and \( A_{\text{exp}} \).

\( A_{\text{prop}} \) is a categorical response, representing the full text of one of five predefined options, as shown in Table~\ref{tab:positivity_negativity_scores}, labeled with scores from “Strongly Negative (-2)” to “Strongly Positive (+2)”. \( A_{\text{exp}} \) provides a natural language explanation based on the selected \( A_{\text{prop}} \), ensuring interpretability and alignment with the chosen action.

The dataset, generated using GPT-4o and Llama-3.1 405B, consists of 10 domains (e.g., healthcare, investment) and 100 binary opposing concepts per domain (e.g., optimism vs pessimism). Each concept is assigned two user traits: \( d_0 \), reflecting "negative" tendencies (e.g., risk aversion, caution), and \( d_1 \), reflecting "positive" tendencies (e.g., proactiveness, autonomy). For each concept, five predefined options are generated with scores from -2 to +2, aligned with \( d_0 \) or \( d_1 \). Contexts (\( c_i \)) are applied to emphasize specific traits:

\begin{spacing}{1.1}
\begin{itemize}[itemsep=1mm, leftmargin=3mm]
    \item \( c_0 \): A baseline context where no specific traits are emphasized.
    \item \( c_1 \): A context that emphasizes unverifiable and negative traits \( d_0 \).
    \item \( c_2 \): A context that emphasizes verifiable and positive traits \( d_1 \).
    \item \( c_3 \): A context with moderate emphasis on \( d_1 \), positioned between \( c_0 \) and \( c_2 \).
\end{itemize}
\end{spacing}

The entire generated test dataset, excluding missing data, consists of 945 instances for GPT-4o and 989 instances for Llama-3.1.

\subsubsection{Analysis of \( A_{\text{prop}} \) Results}

Figure~\ref{fig:fig2} depicts the distribution of \(A_{\text{prop}}\) scores across contexts. Both models exhibited similar tendencies as follows: The baseline (\(c_0\)) skews slightly positive, indicating an inherent LLM bias. \(c_1\) (emphasizing \(d_0\)) shifts the responses negatively, while \(c_2\) (emphasizing \(d_1\)) skews them positively. \(c_3\) (applying weak emphasis on \(d_1\)) results in a moderate positive shift. These findings confirm that adjusting the emphasis on \(d_0\) and \(d_1\) consistently impacts \(A_{\text{prop}}\).

\subsubsection{Analysis of \( A_{\text{exp}} \) Results}

Table~\ref{tab:model_comparison} presents a comparative analysis of \( A_{\text{exp}} \) across different contexts (\( c_i \)), with performance metrics calculated as the average cosine similarity over all trials. Regarding Similarity(\( A_{\text{exp}}, A_{\text{prop}} \)), GPT-4o generally exhibits high similarity values and thus tends to produce explanations consistent with the given proposals. Although Llama-3.1 405B is slightly inferior to GPT-4o in this regard, from Cond0 through Cond3 it shows similar trends in how Similarity(\( A_{\text{exp}}, d_0 \)) and Similarity(\( A_{\text{exp}}, d_1 \)) change. For example, in Cond2 (\(c_{\text{prop}} = c_{\text{exp}} = c_2\)), both models have Similarity(\( A_{\text{exp}}, d_1 \)) surpassing Similarity(\( A_{\text{exp}}, d_0 \)), which can be explained by the fact that \( c_2 \) is a context emphasizing \( d_1 \).

Interestingly, the tendencies diverge between the models in Cond4. In Cond4 (\(c_{\text{prop}} = c_3, c_{\text{exp}} = c_1\)), contradictory contexts (\(c_3, c_1\)) are given to the prompt generating both the proposal and the explanation. Under these conditions, for Similarity(\(A_{\text{exp}}, A_{\text{prop}}\)), GPT-4o’s similarity decreases compared to the baseline (Cond0), whereas Llama-3.1’s similarity increases. In Cond4, we introduce a multi-step reasoning process in order to apply contradictory contexts, which may lead to differing behaviors depending on the model. This indicates that further research is required.

\section{Conclusion}

This study explored privacy preservation in LLM advisors using zero-knowledge proof (ZKP) technologies, particularly zkVMs. Specifically, we demonstrated that zkVMs like RiscZero can transform existing rule-based inference logic into zero-knowledge applications, enabling proof generation and verification within practical timeframes. Furthermore, we introduced a method that uses ZKP to categorize user traits into verifiable segments and validated that the proposed prompt strategies enable LLM advisors to generate consistent proposals and explanations. Future work will explore how users perceive and accept advice generated by these methods.

\begin{table}[tbp]
\centering
\small
\setlength{\tabcolsep}{1pt}
\renewcommand{\arraystretch}{1.1}
\caption{Cosine Similarity of \( A_{\text{exp}} \) Across Contexts (\( c_i \))}
\label{tab:model_comparison}
\resizebox{\columnwidth}{!}{
\begin{threeparttable}
\begin{tabular}{llccccc}
\toprule
\textbf{Model} & \textbf{Criteria} & \makecell[c]{\textbf{Cond0} \\ \(c_{\text{prop}}=c_{\text{exp}}\) \\ \(=c_0\)} & \makecell[c]{\textbf{Cond1} \\ \(c_{\text{prop}}=c_{\text{exp}}\) \\ \(=c_1\)} & \makecell[c]{\textbf{Cond2} \\ \(c_{\text{prop}}=c_{\text{exp}}\) \\ \(=c_2\)} & \makecell[c]{\textbf{Cond3} \\ \(c_{\text{prop}}=c_{\text{exp}}\) \\ \(=c_3\)} & \makecell[c]{\textbf{Cond4} \\ \(c_{\text{prop}}=c_3,\) \\ \(c_{\text{exp}}=c_1\)} \\
\midrule
\multirow{3}{*}{GPT-4o}  & Similarity(\(A_{\text{exp}}, d_0\)) & \large 0.527 & \large 0.581 & \large 0.417 & \large 0.499 & \large 0.526 \\
 & Similarity(\(A_{\text{exp}}, d_1\)) & \large 0.520 & \large 0.438 & \large 0.581 & \large 0.553 & \large 0.532 \\
 & Similarity(\(A_{\text{exp}}, A_{\text{prop}}\)) & \large 0.709 & \large 0.710 & \large 0.715 & \large 0.693 & \large 0.679 \\
\midrule
\multirow{3}{*}{Llama-3.1 405B}  & Similarity(\(A_{\text{exp}}, d_0\)) & \large 0.506 & \large 0.550 & \large 0.410 & \large 0.503 & \large 0.535 \\
 & Similarity(\(A_{\text{exp}}, d_1\)) & \large 0.516 & \large 0.443 & \large 0.600 & \large 0.521 & \large 0.488 \\
 & Similarity(\(A_{\text{exp}}, A_{\text{prop}}\)) & \large 0.661 & \large 0.656 & \large 0.664 & \large 0.633 & \large 0.692 \\
\bottomrule
\end{tabular}
\end{threeparttable}
}
\end{table}


\bibliography{main}

\end{document}